\documentclass[amsmath,amssymb,pre,twocolumn,floatfix]{revtex4}

\def\D{{\rm d}}

\usepackage{color}
\usepackage{graphicx}
\usepackage{dcolumn}
\usepackage{times}
\usepackage{soul}
\usepackage{cleveref}

\begin{document}

\title{Phase separation and folding in swelled nematoelastic films}
\author{A.~P. Zakharov, L.~M. Pismen}
\affiliation{
Technion -- Israel Institute of Technology, Haifa 32000, Israel}

\begin{abstract}
We explore a novel strategy of patterning nematic elastomers that does not require inscribing the texture directly. It is based on varying the dopant concentration that, beside shifting the phase transition point, affects the nematic director field via coupling between the gradients of concentration and nematic order parameter. Rotation of the director around a point dopant source causes topological modification manifesting itself in a change of the number of defects. A variety of shapes, dependent on the dopant distribution, are obtained by anisotropic deformation following the nematic--isotropic transition.
\end{abstract}
 \maketitle

\section{Introduction}
Liquid crystal elastomers, made of cross-linked polymeric chains with embedded mesogenic structures and combining orientational properties of liquid crystals with shear strength of solids,  were envisaged by de Gennes as prototype artificial muscles \cite{degennes97}. Their specific feature is a strong coupling between the director orientation and mechanical deformations  \cite{Warner}, which can be controlled by the various physical and chemical agents. Much attention has been attracted recently to reshaping of nematic elastomers due to anisotropic deformations accompanying phase transition from the nematic to the isotropic state (NIT) {\cite{Warner12,sharon,mostajeran2016encoding}. Similar problems arise in the study of fiber-laden fluid membranes \cite{rey}. A great variety of patterns in thin liquid nematic films is made possible by tangential nematic anchoring on the confining planes. A nematic elastomer film obtained upon polymerisation is further deformed following NIT. 
Most studies concentrate on flat thin films with a prescribed nematic alignment that acquire a certain shape with non-vanishing Gaussian curvature upon NIT \cite{Warner12,sharon,Cirak14,mostajeran2015curvature,mostajeran2016encoding}. In principle, \emph{any} nematic pattern is suitable for creating \emph{some} three-dimensional (3D) shape of a bent shell, though the reverse problem of finding a pattern leading to the desired shape is far harder to solve and its solution is likely, under different circumstances, to be either non-existent or non-unique. 

The Gaussian curvature is rather exceptional among the properties of 3D shells in being uniquely defined, through the famous \emph{Theorema Egregium}, by the surface metric alone. It is, however, not sufficient for visualising the actual shape generated by embedding a surface with a given metric into 3D space, which requires numerical computation \cite{Cirak14,epj15}, except in simplest symmetrical settings yielding surfaces of revolution, such as a cone \cite{Warner12} or a (pseudo)spherical segment \cite{sharon}. The often repeated statement that Gaussian curvature can be created without costs in stretching energy does not apply to the defect cores or boundary layers near interphase boundaries \cite{gemmer2013shape}. 

Some simple nematic patterns can be imposed in a most direct way, without prescribing a particular structure of  tangential anchoring, through boundary conditions on the edges of a finite domain. This, however, leaves very few possibilities open. The equilibrium lowest-energy state in a disk with either normal or tangential anchoring on the edge contains a pair of defects with the charge $+1/2$, while defects with the charge $-1/2$ are obtained in the pretzel topology \cite{epj15}. On a spherical shell, devoid of boundaries, topology requires the total charge equal to two\cite{mermin,kleman}, which is accommodated by four defects with the charge $+1/2$ \cite{Park,nelson}, though their configuration depends on the ratio of splay to bend nematic elasticity \cite{Bowick08prl}.
These patterns lead to a very limited variety of shapes, which all contain singularities at defects locations, smoothed out by bending rigidity of the film. 

A much wider variety can be achieved by prescribing a nematic texture. Experimentalists became adept at imposing a desired orientation on a liquid-crystalline film, either optically \cite{nersisyan2009fabrication,mcconney2013topography} or by surface relief grating \cite{viswanathan1999surface}, \emph{before} it is polymerised into an elastic structure. The various shapes have been produced by heating the frozen texture above the NIT point \cite{Broer14,BroerLa,White,WhiteSM}. \emph{Reversible} transitions of this kind were used for the construction of artificial walkers \cite{Wiersma,japwalk}. Further theoretical models envisage the use of repeated reshaping for propelling nematoelastic walkers  \cite{desimone, pre16} and swimmers \cite{swim}. Other reshaping effects were explored to produce the various bent forms that may serve as actuators \cite{review10,review12,Yusm12,Ionov,review16} and ``4D-printing" of  structures variable in time \cite{print4d}.

We are interested in patterns and shapes created in a more natural way, without inscribing a nematic texture but exploiting spatial variations of the scalar order parameter of the kind that have been induced experimentally via \emph{trans-cis} isomerisation of azobenzene induced by ultraviolet irradiation \cite{Ikeda,Samitsu,Broer}. The direct effect of modification, reflected by the algebraic terms in the Landau--de Gennes energy functional \cite{degennes}, is a change of the scalar nematic order parameter (NOP). In particular, the film can be made to contain both nematic and isotropic domains; in this case, the formation of defects can be avoided, as the energy is lowered when defects drift into an isotropic domain while the circulation of the director field in a nematic domain is conserved. 

The influence of chemical modification may be, however, still more profound due to the coupling between the gradients of concentration and NOP. This interaction has been added in a natural way to the lowest-order free energy functional in several theoretical works \cite{Fukuda,rey04,kopf2013phase} but its role in the formation of patterns and shapes has not been given so far due attention. Although we are not aware of experimental measurements of the respective interaction coefficient, it is expected \emph{a priori} to be comparable with nematic Frank constants. The principal effect of the gradient interaction is \emph{spontaneous anchoring} that governs nematic orientation in transitional domains where both NOP and dopant concentration change \cite{kopf2013phase}. We shall see that this effect may also influence the topology of the nematic field and change the total charge of defects.

In this communication, we explore textures and shapes that can be obtained by taking advantage of a gradual change of dopant concentration, starting with films with tangential nematic orientation and avoiding any externally imposed surface patterning or anchoring at the edges. Although a wide variety of  scalar NOP patterns can be brought about by optically induced isomerisation, we concentrate on distributions that would be obtained chemically by a signalling species changing the dopant concentration. This may mimic biological signalling  \cite{Wolpert} modifying the properties of heretofore undifferentiated cells in the epithelial tissue of an embryo and thereby initiating complex reshaping necessary for morphogenetic development. This approach opens new ways of \emph{biomimetic} reshaping, that can be further developed to include multiple morphogenetic gradients \cite{pre11} and reshaping due to modification of the various mechanical properties \cite{du}.

\section{Basic equations}

Assuming planar orientation of the nematic director constant across the film thickness, the 2D nematic order parameter is presented in Cartesian coordinates as a traceless symmetric tensor
\begin{equation}
\mathbf{Q} = \left(\begin{array}{cc}p & q \\ q & -p \end{array} \right)
= \frac {S}{\sqrt{2 }}\,\left(\begin{array}{cc}\cos 2\theta & \sin 2\theta \\ \sin 2\theta & -\cos 2\theta \end{array} \right),
\end{equation} 
where $S=\sqrt{\mathrm{Tr}(\mathbf{Q}\cdot\mathbf{Q})}=\sqrt{p^2+q^2}$ is the scalar NOP, and $\theta$ is the director orientation angle. The nematic energy per unit thickness is expressed as ${\cal F}_n = \int {\cal L}_n\D^2 \mathbf{x} $ with the Lagrangian
\begin{align}
{\cal L}_n = & -\frac{1 - \alpha c}{2}Q_{ij}Q_{ij} +
\frac 14\left(Q_{ij}Q_{ij} \right)^2 - \beta \nabla_i c \nabla_j
Q_{ij} \notag \\
& + \frac{\kappa_1}{2} \left|\nabla_i Q_{ij} \right|^2 + \frac{\kappa_2}{4}
\sum_{ijk}\left(\nabla_iQ_{jk}\right)^2. \label{eq:LQdef}
\end{align}
The linear dependence of the coefficient in the first term on the dopant concentration $c$ with the slope $-\alpha$ implies that the isotropic state prevails at high $c$, and NIT takes place in a uniform material at $c=c^*=1/ \alpha$. The term of the 4th order in \textbf{Q} is standard; the cubic term vanishes identically in 2D. The third term, essential for the following analysis, describes the lowest-order interaction between inhomogeneities in composition and NOP fields allowed by symmetry \cite{Fukuda,rey04,kopf2013phase}. The last two distortion energy terms present a simplification of the full Ericksen's formula \cite{Ericksen}. Similar to the one-constant approximation, they ignore orientation dependence of elastic coefficients but, unlike the commonly used Frank energy expressed through a unit vector \cite{degennes}, take into account changes of the scalar NOP.  The Lagrangian is scaled by a characteristic nematic transition energy per unit area $E_0$ to bring the maximum value of $S$ at $c=0$ to unity. The constants $\kappa_i$ defining the squared healing length are positive, while $\beta$ may have either sign. Summation over repeated indices is presumed. 

The role of the gradient interaction term is made transparent when only the scalar NOP $S$ changes along the dopant concentration gradient, taken as the $x$-axis, while the nematic orientation angle $\theta$ remains constant. In this case, this term reduces to 
\begin{align}
 - 2^{-1/2}\beta (c_xS_x \cos 2\theta + c_x S_y\sin2\theta).
 \label{LnS}
\end{align}
Assuming that $S$ decreases with $c$ ($\alpha>0$), the optimal angle, reducing the overall anergy at $\beta<0$,  is $\theta=0$, so that the director tends to orient along the concentration gradient. At $\beta>0$, on the opposite, the lowest energy is attained at $\theta=\pi/2$, when the director is oriented normally to the gradient (along the $y$-axis); the change of  $S$ along the $y$-axis does not affect this argument. This term is therefore responsible for spontaneous anchoring at a nematic-isotropic interface\cite{kopf2013phase}. 

\begin{figure}[t]
\centering
\includegraphics[width=.3\textwidth]{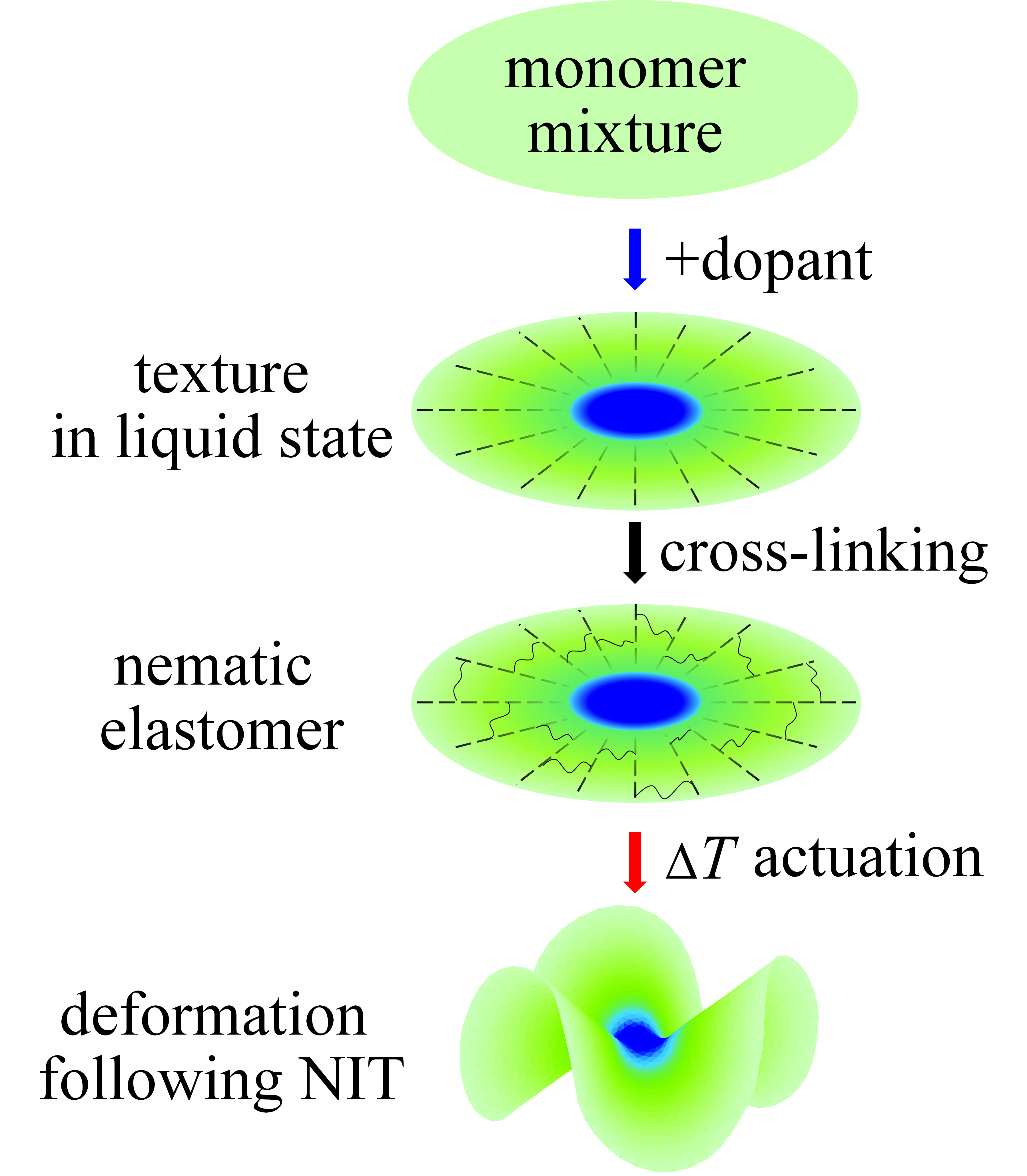} 
\caption{A scheme of the film preparation and actuation. The darker (blue) shading marks the isotropic domain.}
\label{Scheme}
\end{figure}

In the following, we simulate a usual experimental procedure whereupon nematic alignment is established in the liquid state before polymerisation, and then the elastomer is heated to bring it to the isotropic state and cause it to deform (see Fig.~\ref{Scheme}). The difference from the standard procedure is in establishing a controlled dopant distribution at the first stage. Accordingly, we start with defining nematic alignment in a flat film minimising the energy functional defined by Eq.~\eqref{eq:LQdef} in an imposed concentration field with either constant or variable parameter $\beta$. It is assumed that a change in dopant concentration (unlike swelling by a solvent) does not cause a volume change. The computed distribution of NOP defines the local intrinsic deformation tensor $\overline{\mathbf{u}}$ following NIT. The resulting 3D shape is obtained by minimising the elastic deformation energy $\mathcal{F}_e= \int {\mathcal L}_e\D^2 \mathbf{x} $, where the Lagrangian ${\mathcal L}_e$, scaled by the ratio of the effective shear modulus $\mu h$ (where $h$ is the film thickness) to $E_0$, is defined in the thin film approximation as \cite{epj15}   
\begin{equation}
\mathcal{L}_e = \frac 12 \left[ |\mathbf{u}-\overline{\mathbf{u}}|^2
 +  \frac { h^2}{9} \mathrm{Tr}( \mathbf{C}^2  )  \right] .
 \label{ElastEnergy}
\end{equation}
The first term, containing the deformation tensor $\mathbf{u}$, defines the elastic part of in-shell strain, while the second term containing the trace of the squared curvature tensor $ \mathbf{C}$ is due to flexural rigidity of a bent shell;  the numerical coefficient corresponds to an incompressible material. 

Interaction between the nematic order and elasticity is expressed in Eq.~\eqref{ElastEnergy}  in an implicit way by the intrinsic deformation tensor $\overline{\mathbf{u}}$ determined by the anisotropic extension and contraction of the elastomer due to changing NOP.  In a coordinate framework aligned with the director, the material contracts upon NIT along the director by a factor $\lambda$ dependent on the change of $S$, and extends normally to the director by the factor $\lambda^{1/2}$ to preserve the volume\cite{Warner}; this causes both  in-shell extension and an increase of the shell thickness. A more complicated procedure\cite{kopf2013phase}, which includes splitting the total deformation tensor into the elastic and intrinsic parts, with the latter expressed by an unisotropic ``growth tensor" reduces to the same formula in the approximation neglecting higher-order corrections to the nematic energy \eqref{eq:LQdef} caused by deformation.

Reshaping of thin shells is dominated by bending rigidity, whereas in-shell deformations are strongly discouraged. If they are assumed to be totally excluded, the problem reduces to constructing a surface with a metric determined by intrinsic deformation\cite{Warner12,Cirak14,epj15}. In this approximation, deformation of a nematoelastic shell is similar to that of membranes governed by curvature-elasticity theory \cite{Helfrich}. In application to nematic membranes\cite{Park,Biscari}, the bending energy is proportional to the nematic, rather than mechanical elasticity. This component is present also in nematoelastic shells but mechanical elasticity is dominating here. Our computation algorithm retains both extensional and bending terms, which allows us to explore the effect of varying shell thickness.

The energy functionals $\mathcal{F}_n$, $\mathcal{F}_e$ are discretised on a domain  triangulated by the Delaunay algorithm \cite{Delaunay}, with the NOP, concentration, and deformation fields defined at the mesh nodes. The concentration field on a flat film due to a source located either on a domain boundary or at an isolated internal point is determined either analytically (in symmetric configurations) or numerically. The scalar NOP is determined algebraically, neglecting nematic elasticity, as $S=\sqrt{1-\alpha c}$. An exception is the boundary layer near the locus $S=0$ (see below).  
In symmetric configurations, the director field on a flat film may be determined straightforwardly by inspecting the gradient interaction term in Eq.~\eqref{eq:LQdef}. 
Otherwise, it is found numerically by minimising the nematic energy $\mathcal{F}_n$. The minimum energy configuration is obtained
by evolving the orientation angle $\theta$ in pseudo-time $t$ as $\partial\theta/\partial t=-\delta \mathcal{F}_n / \delta \theta$, continued until an equilibrium state is reached. 

At the next stage, the shapes formed following NIT are computed by minimising the elastic energy $\mathcal{F}_e$ discretised as 
\begin{equation}
\mathcal{F}_e = \frac 12 \sum_\mathrm{nodes} \left[  \frac{h_i}{2} \sum_\mathrm{adj.n} \left(\frac{l_{ij} }{ \overline{l}_{ij}} -1\right)^2 
+ \frac{h_i^3}{9} \sum_\mathrm{adj.t} \left\langle \frac{1-\mathbf{m}_i\cdot \mathbf{m}_{ij}}{l_{ij}^2}\right\rangle_j  \right].
 \label{eq:MC}
\end{equation}
The first term accounts for a deviation of the observed length $l_{ij}$ of an edge between adjacent nodes $i$ and $j$ from its ``optimal" length $\overline{l}_{ij}$ due to intrinsic elongation or shortening following NIT.  If the length of an edge parallel to the director shortens by a factor $\lambda$, the edge normal to the director, as well as the shell thickness, elongate by the factor $\lambda^{1/2}$ to preserve the volume. The respective length transformation matrix for an edge at an angle $\psi$ to the director is $\mathbf{R}^{-1}(\psi)\boldsymbol{\Lambda}\mathbf{R}(\psi)$, where $\boldsymbol{\Lambda}$ is the diagonal matrix with the elements $\{1/\lambda, \sqrt{\lambda} \}$ and $\mathbf{R}(\psi)$ is the 2D rotation matrix. This yields, for an edge with the original of length $l_0$,
\begin{equation}
\frac{\overline{l}}{l_0} =\sqrt{\frac 14\left(\sqrt{\lambda }-\frac{1}{\lambda}\right)^2  \sin^2 2\psi + \left(\sqrt{\lambda }\sin ^2 \psi+\frac{\cos^2 \psi}{\lambda} \right) ^2}.
\label{length}
\end{equation}
In the second term of Eq.~\eqref{eq:MC}, $\mathbf{m}_{ij}$ denotes the normal to a $j$th tile of those surrounding an $i$th node, and $l_{ij}$ is the distance between the node and the tile centre. The average over all neighbouring tiles measures the deviation from their average orientation $\mathbf{m}_{i}$, which accounts for the curvature of the shell. Similar to minimising the nematic energy, the node coordinates $\mathbf{x}_i$ are evolved in pseudo-time following the evolution equation $\partial \mathbf{x}_i/\partial t=-\delta \mathcal{F}_e / \delta \mathbf{x}_i$.

\section{Reshaping a disk}

\begin{figure*}[t]
	\centering
\begin{tabular}{cccc}
(a) & (b)  \\
\includegraphics[width=.20\textwidth]{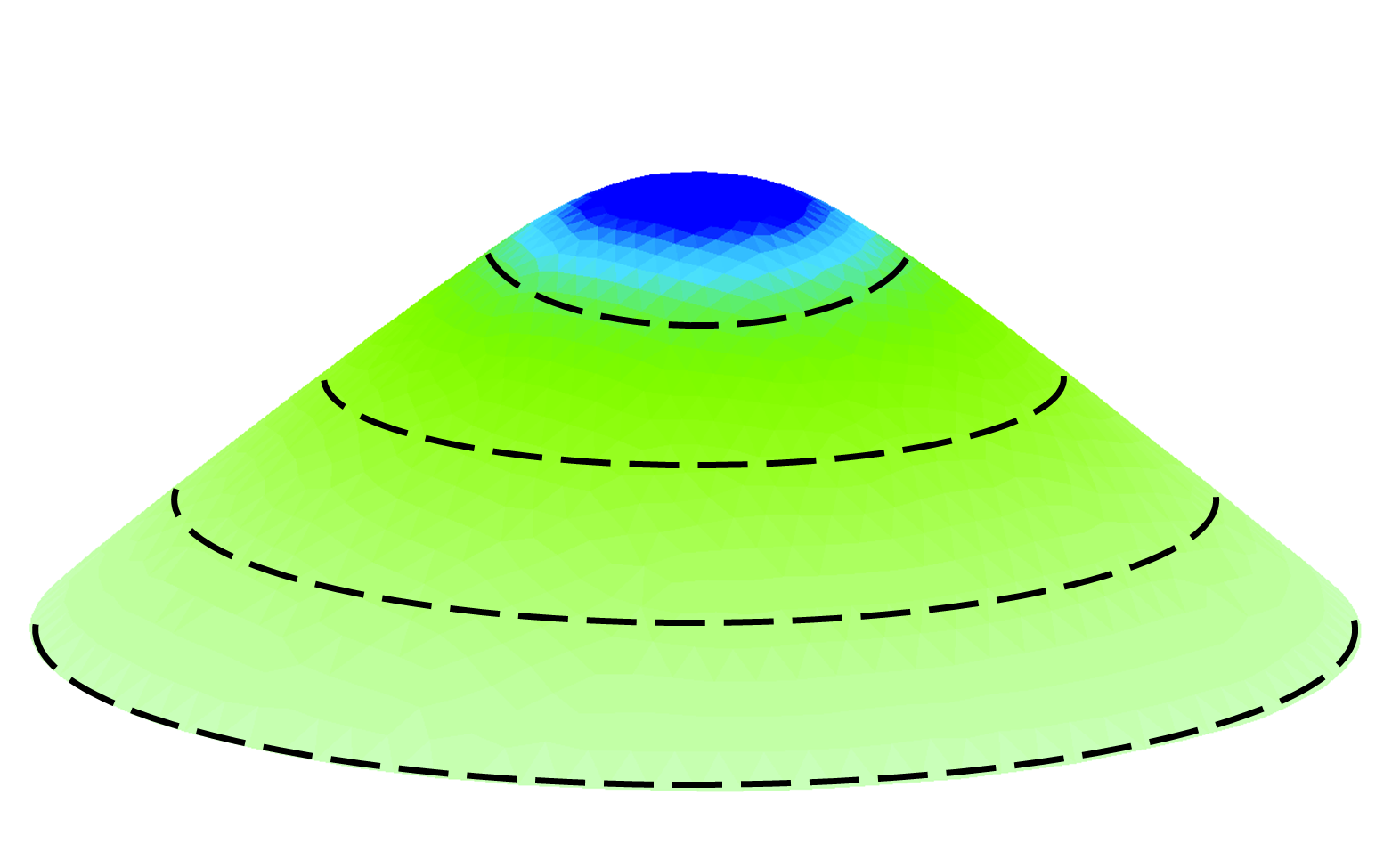} &
\includegraphics[width=.68\textwidth]{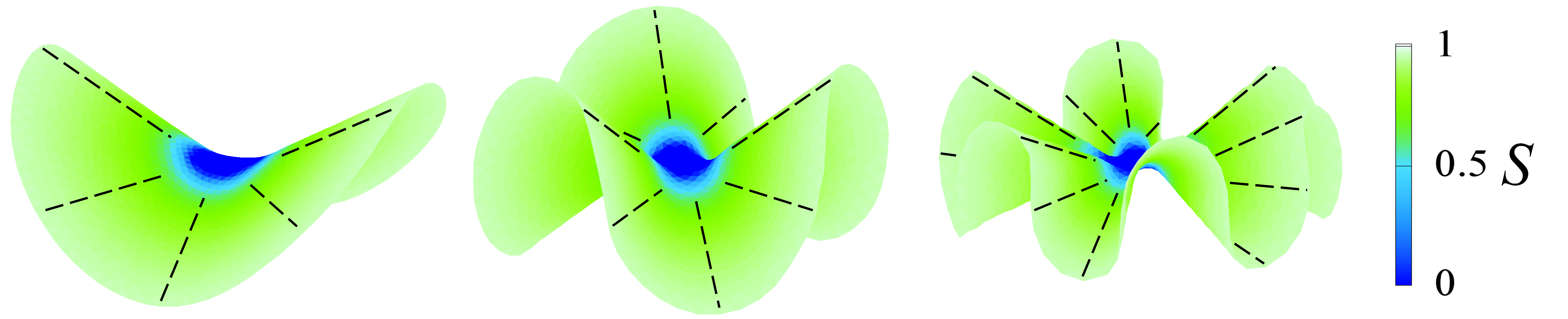} 
\end{tabular}
\caption{A graded cone at $\beta=1$ (a) and anticones with different symmetries ($n=2,3,6$) at $\beta=-1$  (b) formed \emph{following} NIT. The darker (blue) shading corresponds to lower $S$ and dashes mark the director orientation \emph{before} NIT.}
\label{Circle}
\end{figure*}

The simplest example is a disk of radius $L$ with a radial concentration distribution, leading to $S$ dependent on the radial coordinate $r$ only. We use a concentration distribution obeying the decay-diffusion equation 
\begin{equation}
 \frac{1}{r}\frac{\D}{\D r}\left( r\frac{\D c}{\D r} \right)- k^2 c(r)=0,
  \label{eq:dif}
\end{equation}
where $k$ is the inverse diffusional length (further taken as the length scale). The solution due to a source of unit intensity at the origin  is $c=(2\pi)^{-1}K_0(kr)$, where $K_0(x)$ is a modified Bessel function.

With no edge anchoring, the natural boundary condition for the orientation angle $\theta$ is $\theta'(r)=0$ at $r=L$, so that a symmetric a nematic alignment field $\theta=2(\phi+\phi_0)$ depends on the angular coordinate $\phi$ only, $\phi_0$ being a constant phase shift. Then Eq.~\eqref{eq:LQdef} reduces in polar coordinates to 
\begin{equation}
{\cal L}_n =  -\frac{1 - \alpha c}{2}S^2 +
\frac 14 S^4 - \frac{\beta}{\sqrt{2}} c'(r) S'(r) \cos 2\phi_0 + \frac {\kappa}{4} S'(r)^2,
 \label{eq:LQS}
\end{equation}
where $\kappa=\kappa_1 + \kappa_2$.

The gradient interaction term is minimised by choosing $\cos 2\phi_0$ equal to 1 or $-1$, respectively, at negative or positive $\beta$, in agreement with the discussion following Eq.~\eqref{LnS}. The first case, with $\phi_0=0$, corresponds to the radial (aster), and the second, with $\phi_0=\pi/2$, to the circumferential (vortex) alignment. 
In the last case, the shape emerging following NIT retains circular symmetry, and has a form of a ``graded cone" formed by shortening of the circumference of each circle with the radius $r$ by the factor $\lambda=1+a S(r)$, where $a>0$ is the expansion coefficient, and extending the radius and the thickness by the factor $\lambda^{1/2}$ to preserve the volume (Fig.~\ref{Circle}a). With the radial dependence of $S(r)$ excluded, this reduces to a standard cone generated by the vortex structure in the vicinity of a defect of unit charge \cite{Warner12} (which tends to split into a pair of half-charged defects, unless inscribed and frozen). No defects are formed, however, when the material remains isotropic near the centre and the top of the graded cone remains flat. 

 Using the cylindrical coordinates $x^1=\rho,\, x^2=\phi,\, x^3=z$ with the Euclidean metric $g_{11}=g_{33}=1, \; g_{22}=\rho^2$, we obtain \cite{lp14} for a target surface with the elevation $z(r)$ and the local radius $\rho(r)$ the diagonal metric tensor with the elements
\begin{equation}
\gamma_{11}= z'(r)^2+\rho'(r)^2=\lambda , \quad \gamma_{22}=\rho^2=(r/\lambda)^2.
  \label{eq:g}
\end{equation}
Using the last expression, $z(r)$ is integrated to
\begin{equation}
z(r) = \int \sqrt{ 1+aS(r) -
\left[ \frac{\D}{\D r}\left( \frac{r}{1+aS(r)} \right)\right]^{2}}\D r.
  \label{eq:hr}
\end{equation}

\begin{figure}[t]
\centering
\includegraphics[width=.48\textwidth]{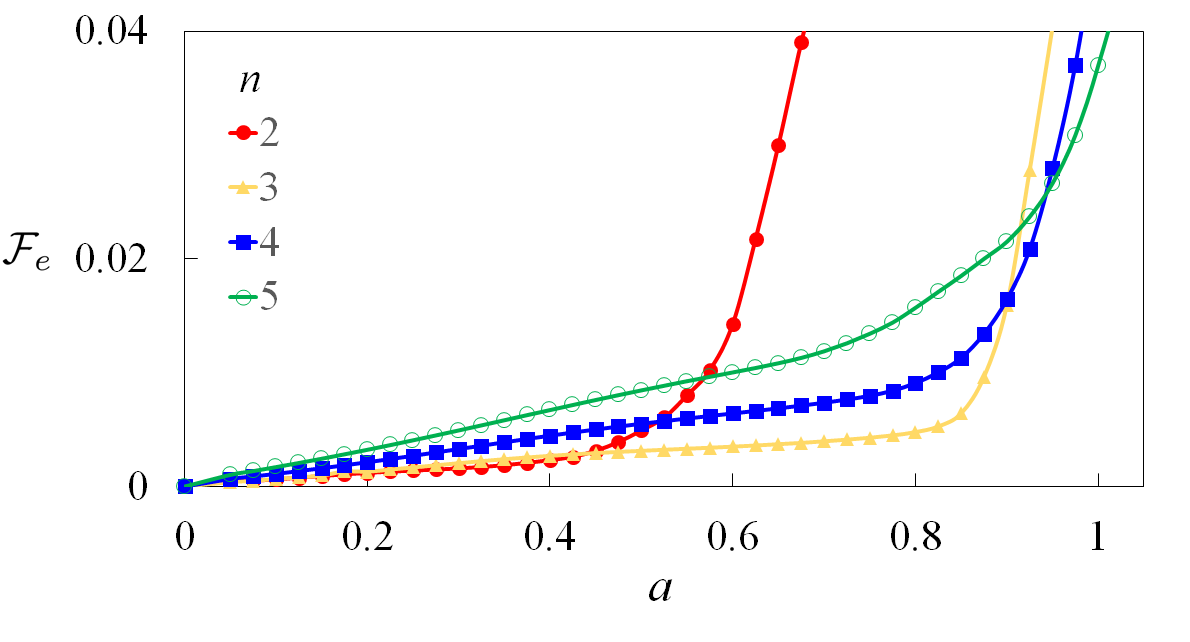} 
\caption{Dependence of  elastic energy on the extension coefficient $a$ for shapes with the different number of  ``petals" $n$. Parameters: $h=0.01, \,\alpha=5, \,\kappa_1=\kappa_2=1, \,L=2$. }
\label{CircleF}
\end{figure}

\begin{figure}[t]
\centering
\includegraphics[width=.48\textwidth]{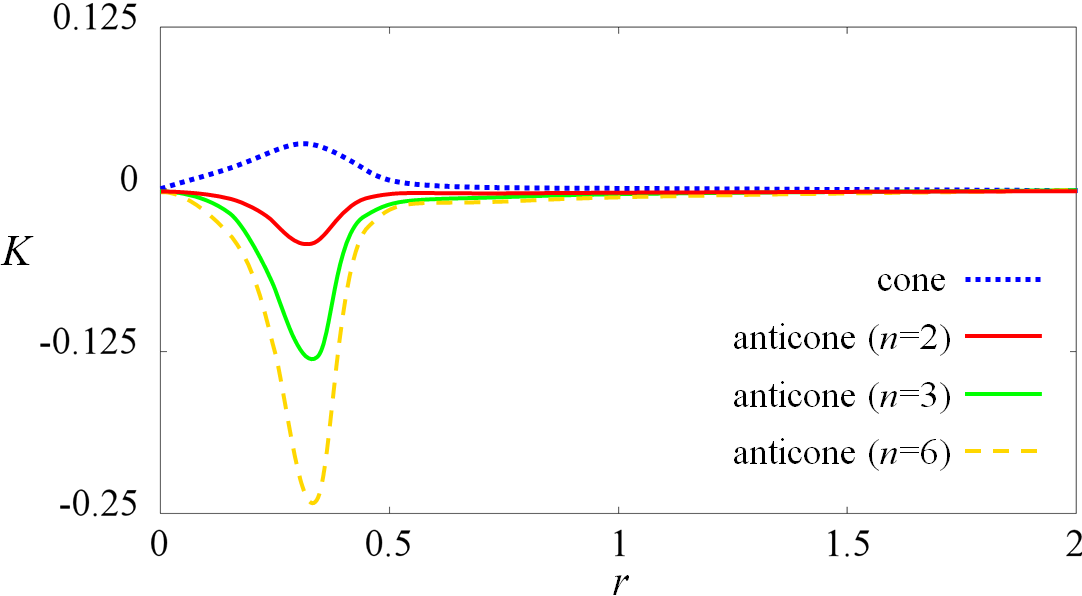} 
\caption{Radial dependence of  the Gaussian curvature for a graded cone (upper curve) and anticone (lower curve) for $\alpha=5, \,a=0.2$.}
\label{gauss}
\end{figure}

If nematic elasticity is neglected, $S=\sqrt{1-\alpha c}$ is defined by the algebraic energy terms only. Suppose $c(r_0)> 1/\alpha$ within the circle $r<r_0$, so that $S$ vanishes there. Just outside the isotropic circle, at $r=r_0+\epsilon x$, $S \sim \sqrt {\epsilon x}$, and, since $S'(r) \sim 1/ \sqrt {\epsilon x}$, elastic terms cannot be neglected in the boundary layer with the width $\epsilon \sim \sqrt {\kappa} \ll 1$ comparable with the healing length. Within this layer, retaining the leading energy terms and approximating the concentration by the linear function $c(x)=-qx$, the equation of $S(x)$ obtained by varying Eq.~\eqref{eq:LQS} reduces to
\begin{equation}
\kappa S''(x) + qx S=0.
  \label{eq:difs}
\end{equation}
The solution vanishing at $x=0$ is expressed through the Airy functions Ai$(x)$, Bi$(x)$:
\begin{equation}
S(x) = s_0 \left[ \sqrt{3}\mathrm{Ai}\left(-(q/\kappa)^{1/3}x\right)
- \mathrm{Bi}\left(-(q/\kappa)^{1/3}x\right)\right],
  \label{eq:sols}
\end{equation}
where $s_0$ is a constant to be obtained by matching to the outer algebraic solution. The derivative of this function at the circle where $S$ vanishes, $S'(0)=2(q/3\kappa)^{1/3}/\Gamma(1/3)$ (where $\Gamma$ is the gamma function) is finite, and therefore a singularity is avoided. At $x=r-r_0 \ll 1$, the last term under the radical in Eq.~\eqref{eq:hr} is of $O(x^2)$ and can be neglected, leading to the elevation over the flat inner circle $z \approx x$ at $x \to 0$. The jump of the slope $z'(r)$ is smoothed out by stretching near the circle $S=0$ \cite{gemmer2013shape}, and no singularity arises in a numerical solution (see Fig.~\ref{Circle}a).

At $\beta<0$, the preferential alignment is aster, so that the radius of the disk should contract and its circumference expand following NIT. In this case, the emerging shape loses circular symmetry \cite{Warner12}. In order to obtain forms with a desired number of  ``petals" $n$ (Fig.~\ref{Circle}b), we started computations from preassigned forms with the desired symmetry, setting tentatively $z=f(r)\cos n\phi$ with a smooth function $f(r)$ vanishing at the origin. At small $a$, forms with different $n$ coexist but the lowest value $n=2$ yields the lowest energy, the others being metastable equilibria. As $a$ grows, shapes with higher $n$ become preferable, while the energy of forms with lower $n$ sharply increases near their existence limit (Fig.~\ref{CircleF}).

Using the induced orthogonal metric $g_{ii}=b_i^2$ with $b_1=\sqrt{\lambda}, \, b_2=r/\lambda$, the Gaussian curvature of a graded cone is computed in a standard way as $K(r)=-(b_1b_2)^{-1}\D_r[b_1^{-1}b_2'(r)]$. For an anticone, we compute the Gaussian curvature of forms with different symmetry numerically. 
Unlike Ref.~\cite{Warner12} where Gaussian curvature is localised at distances from the defect core comparable with the film thickness where stretching and bending energies equalise, it is distributed here over a large range of $O(k^{-1})$ due to a gradual change of $S(r)$ (Fig.~\ref{gauss}). For a graded cone,  $K(r)$ is positive, while for an anticone it is negative everywhere. 

\section{Topology change}

\begin{figure}
	\centering
\begin{tabular}{cc}
(a) & (b) \\
\includegraphics[width=.24\textwidth]{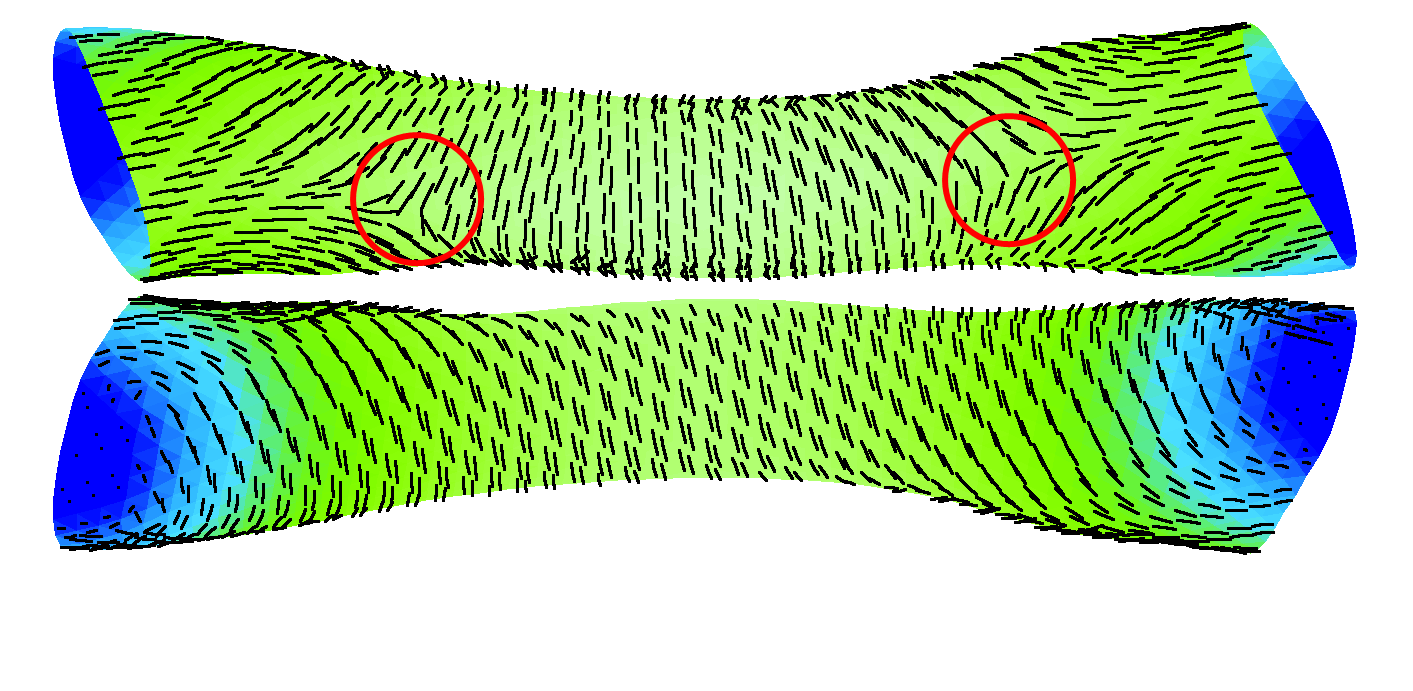}
& \includegraphics[width=.24\textwidth]{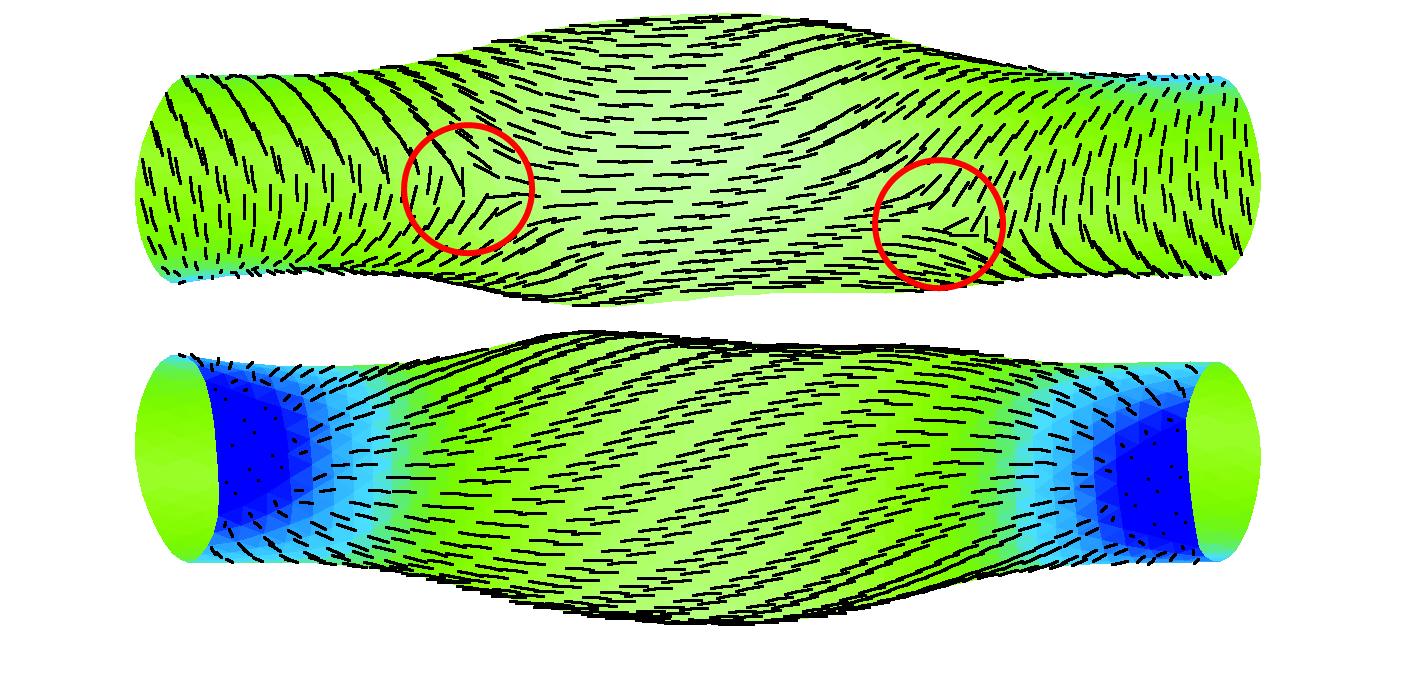}\\
(c) &(d)  \\
\includegraphics[width=.24\textwidth]{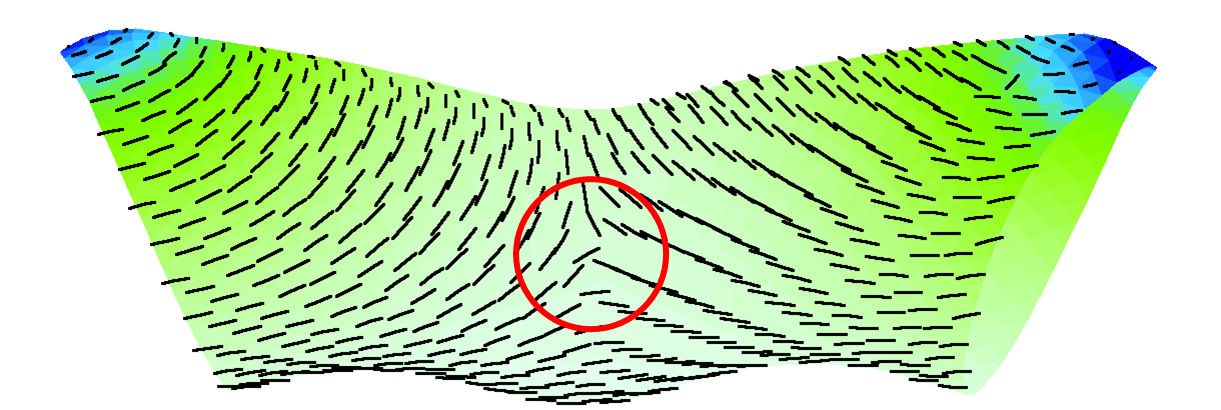} &
\includegraphics[width=.24\textwidth]{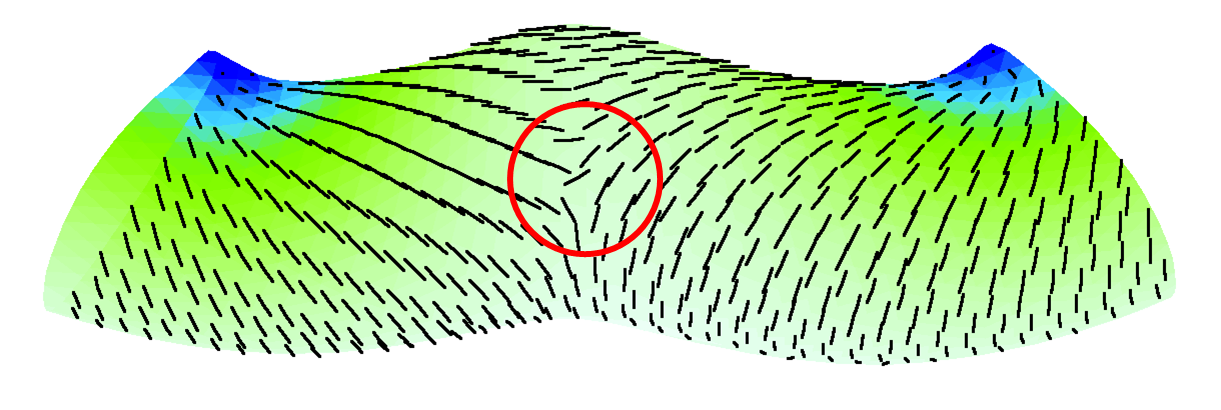} \\
(e) & (f) \\
\includegraphics[width=.24\textwidth]{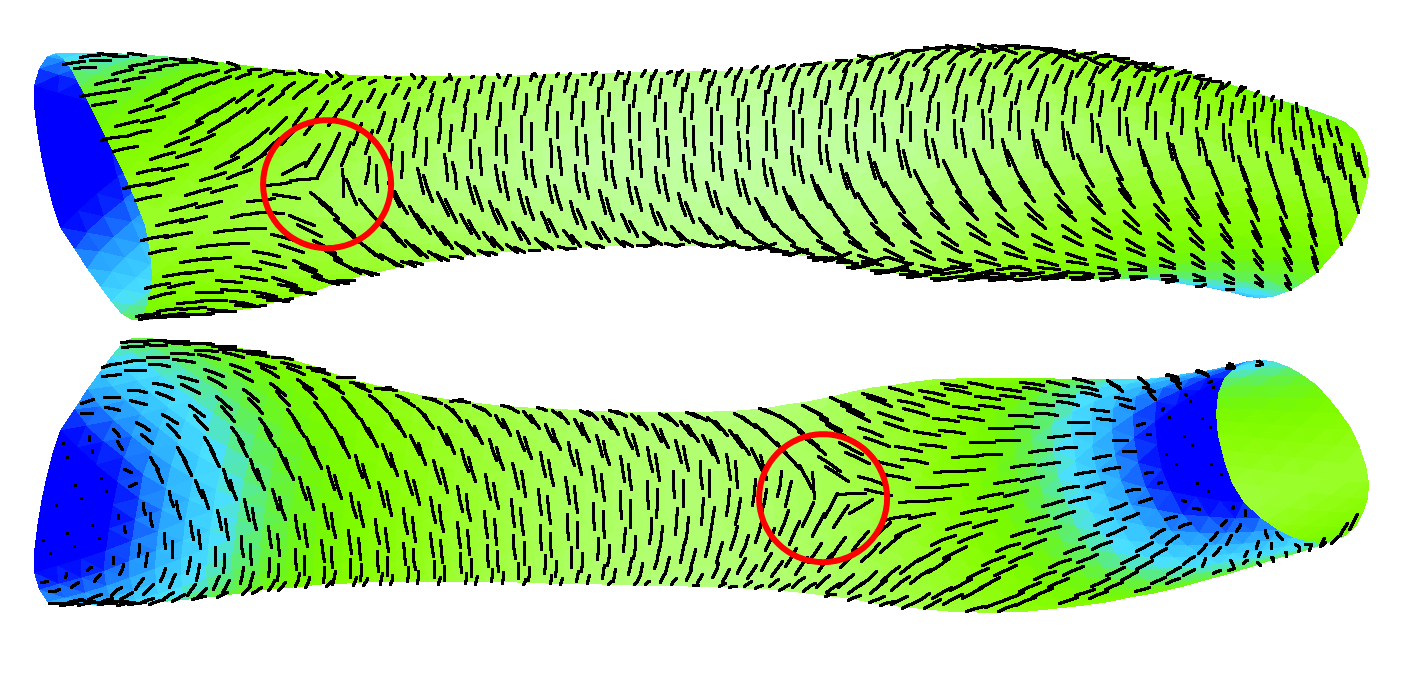}
& \includegraphics[width=.24\textwidth]{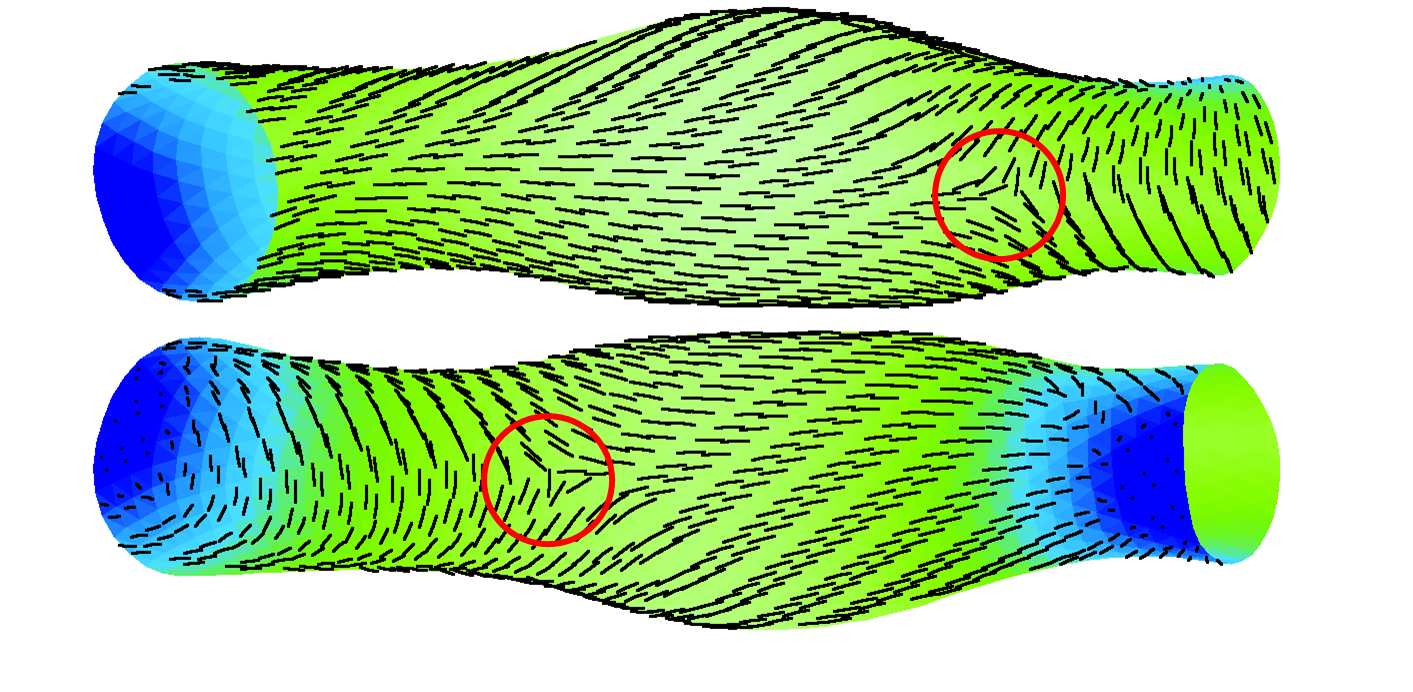}\\
\end{tabular}
\caption{Deformed cylinders with two point concentration sources. Pairs of views, rotated about the cylinder axis by $\pi$, are shown (except in c,d). The darker (blue) shading corresponds to lower $S$ and dashes mark the director orientation before NIT, with locations of defects marked by circles. Parameters: $a=0.2$ (a--f), $L=5, \, R=0.5$. (a,b,e,f), $L=2\pi \, R=1$ (c,d), $\beta=1$ (a,c),  $\beta=-1$ (b,d), $\beta$ changing axially from $+1$ to $-1$ (e,f). Other parameters as in Fig.~\ref{CircleF}.}\label{Tube}
\end{figure}

\begin{figure}[t]
	\centering
\includegraphics[width=.48\textwidth]{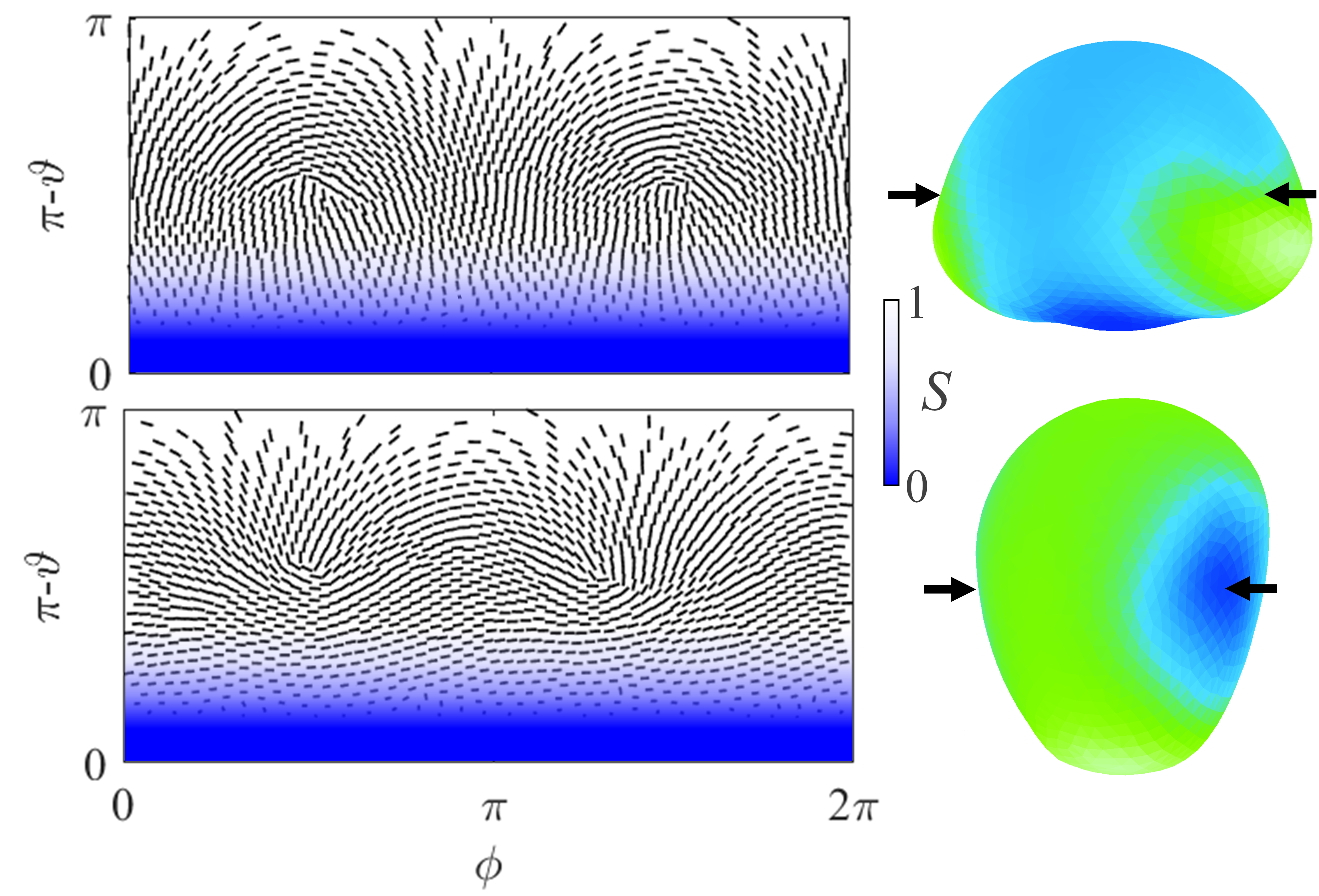} 
\caption{Left: the nematic texture on a sphere of unit radius (shown in the Mercator projection) for $\beta=-1$ (above) and $\beta=1$ (below). The darker (blue) shading corresponds to lower $S$ and dashes mark the director orientation before NIT. Right: respective shapes following NIT with shading showing the local radius. The dopant source is at the south pole, and the defects locations (before NIT) are shown by arrows. Parameters are as in Fig.~\ref{CircleF}.}\label{sphere}
\end{figure}

The nematic alignment field on a cylinder with a uniform concentration source at the open ends is smooth, with the director aligned either axially or circumferentially, dependent on the sign of $\beta$. With $\beta$ varying along the axis (which may be caused by varying concentration of another dopant), the director rotates gradually, and no defects arise. The shapes following NIT retain axial symmetry, with the radius bulging or shrinking, according to the local alignment, which is constant along the circumference.

More interesting textures are induced by \emph{point} dopant sources. The gradient interaction term brings about, independently of the sign of $\beta$, rotation of the director by $\pi$ around a point source located on the boundary, which effectively changes the topology of the nematic alignment field. With symmetric sources at both ends, the combined rotation by $2\pi$ should be compensated by either a pair of defects with the charge $-1/2$ or a single defect with the charge $-1$. The former possibility is preferable, in view of the lower defect energy, and is indeed obtained by minimising the nematic energy, as seen in Fig.~\ref{Tube}. Textures with defects are specific to a cylinder, and defects disappear if the cylinder is cut into a square sheet lacking the periodicity along the circumference. 

The defects are placed axially on the side opposite to the sources in a long cylinder
 (Fig.~\ref{Tube}a,b) but are shifted to a middle location with circumferential separation when the cylinder is squat, with the circumference $2\pi R$ comparable to the length $L$ (Fig.~\ref{Tube}c,d). The textures shown in  Fig.~\ref{Tube}a,c correspond to $\beta=1$. The sign reversal just causes the entire distribution to rotate by $\pi/2$ but the shapes following NIT (Fig.~\ref{Tube}b,d) are, of course, very much different, with radial bulging and shrinking interchanged. 
When $\beta$ varies axially, changing from $+1$ to $-1$ along the length of the cylinder, the defects are located on the opposite sides. In this case, there are two alternative configurations with equal nematic energy but leading to different shapes (Fig.~\ref{Tube}e,f), and the asymmetry is more pronounced. The deformation at defects is not resolved here, and a trefoil pattern \cite{epj15} should be observed at higher resolution when the sheet thickness is comparable with the healing length. In all cases, the formation of defects is necessitated topologically by circumferential periodicity, and if a cylinder is cut to a strip, defect-free textures become possible.       

The dopant concentration distribution on a spherical surface induced by a unit point source at the south pole, obtained by integrating the diffusion-decay equation in spherical coordinates, is expressed through a Legendre function $P_{n}(\cos \vartheta )$ with the index $n=\frac 12-\sqrt{\frac14-(kR)^2}$, where $\vartheta$ is the polar angle and $R$ is the radius of the sphere. The flux at the source is reduced to unity by multiplying this function by the normalising factor $\Gamma(n)\Gamma(n')/4\pi$, where $n'=\frac 12+\sqrt{\frac14-(kR)^2}$. 

The director rotates by $2\pi$ around the source, and therefore it is sufficient to have two, rather than four, defects with the charge $+1/2$ to ensure the required topological charge equal to two. Indeed, the texture minimising the nematic energy, shown in Fig.~\ref{sphere}, displays two $+1/2$ defects placed symmetrically at mutual longitudinal separation equal to $\pi$; the latitudinal position depends on the dimensionless radius $kR$. The textures at $\beta=\pm1$ are rotated by $\pi/2$ relative one to the other and lead to distinct (oblong or oblate) shapes.

\section{Conclusion}
The shapes presented above are just a small sample of the great variety that can be produced by varying the dopant concentration that can be carried out, at the present, most readily by optically induced isomerisation, without inscribing the texture directly. Besides modifying the scalar NOP, dopant gradients affect the nematic alignment in transitional zones through coupling terms in the nematic energy functional. The most important qualitative effect is \emph{topological} modification, which allows one to change the number of defects by adding circulation around a point source of the dopant or absorbing defects in an isotropic domain. 

\emph{Acknowledgement} This research is supported by Israel Science Foundation  (grant 669/14).

\end{document}